\documentclass{sf2a-conf2013}
\usepackage{graphicx}
\usepackage{hyperref}
\usepackage[]{natbib}  
\usepackage{epstopdf}

\def\BibTeX{{\rm B\kern-.05em{\sc i\kern-.025em b}\kern-.08em
    T\kern-.1667em\lower.7ex\hbox{E}\kern-.125emX}}
\bibpunct{(}{)}{;}{a}{}{,}  

\newcommand{\ath}{{\it Athena+}}
\newcommand{\xmm}{{\it XMM-Newton}}
\newcommand{\chan}{{\it Chandra}}

\begin{document}

\TitreGlobal{SF2A 2013}


\title{Athena+:  The first Deep Universe X-ray Observatory}

\runningtitle{The \ath\ X-ray Observatory}
\author{D. Barret}\address{$^{a}$Universit\'e de Toulouse; UPS-OMP; IRAP; Toulouse, France \& $^{b}$CNRS; Institut de Recherche en Astrophysique et Plan\'etologie; 9 Av. colonel Roche, BP 44346, F-31028 Toulouse cedex 4, France}
\author{K. Nandra}\address{Max-Planck-Institut fur extraterrestrische Physik, P.O. Box 1312, Giessenbachstrasse, 85741 Garching, Germany}
\author{X. Barcons}\address{Instituto de F'sica de Cantabria, Edificio Juan Jord\'a, Avenida de los Castros, s/n, E-39005 Santander, Cantabria, Spain}
\author{A. Fabian}\address{Institute of Astronomy, University of Cambridge, Madingley Road, Cambridge, CB3 0HA, United Kingdom}
\author{J.W. den Herder}\address{SRON Netherlands Institute for Space Research,
Sorbonnelaan 2,
3584 CA Utrecht,
Netherlands}
\author{L. Piro}\address{INAF,
IASF Sezione di Roma,
Via Fosso del Cavaliere 100,
Tor Vergata,
00133 Roma,
Italy}
\author{M. Watson}\address{University of Leicester,
Department of Physics \& Astronomy, 
University Rd,
Leicester LE1 7RH,
United Kingdom}
\author{J. Aird}\address{Department of Physics,
Durham University,
South Road,
Durham,
DH1 3LE,
UK}
\author{G. Branduardi-Raymont }\address{University College London,
Mullard Space Science Laboratory,
Holmbury St Mary,
Dorking RH5 6NT,
United Kingdom}
\author{M. Cappi}\address{INAF,
IASF Bologna,
Via P Gobetti 101,
40129 Bologna,
Italy}
\author{F. Carrera}\address{Instituto de Fisica de Cantabria (CSIC-UC),
Avda. de los Castros,
39005 Santander, Spain}
\author{A. Comastri}\address{INAF
Osservatorio Astronomico di Bologna,
Via Ranzani 1,
40127 Bologna,
Italy}
\author{E. Costantini}\address{SRON, Netherlands Institute for Space Research, Sorbonnelaan 2, 3584 CA Utrecht, Netherlands}
\author{J. Croston}\address{University of Southampton,
School of Physics and Astronomy,
Highfield Campus,
Southampton SO17 1BJ,
United Kingdom}
\author{A. Decourchelle}\address{CEA Saclay,
Service d'Astrophysique,
L'Orme des Merisiers Bat 709,
BP 2,
91191 Gif-sur-Yvette Cedex,
France}
\author{C. Done}\address{Department of Physics,
Durham University,
South Road,
Durham,
DH1 3LE,
UK}
\author{M. Dovciak}\address{Astronomical Institute AS CR, Fricova 298, CZ-25165 Ondrejov, Czech Republic}
\author{S. Ettori}\address{INAF,
Osservatorio Astronomico di Bologna,
Via Ranzani 1,
40127 Bologna,
Italy
}
\author{A. Finoguenov}\address{Department of Physics, University of Helsinki, Gustaf Hallstromin katu 2a, 00014 Helsinki, Finland}
\author{A. Georgakakis}\address{Max-Planck-Institut fur extraterrestrische Physik, P.O. Box 1312, Giessenbachstrasse, 85741 Garching, Germany}
\author{P. Jonker}\address{SRON Netherlands Institute for Space Research,
Sorbonnelaan 2,
3584 CA Utrecht,
Netherlands}
\author{J. Kaastra}\address{SRON Netherlands Institute for Space Research,
Sorbonnelaan 2,
3584 CA Utrecht,
Netherlands}
\author{G. Matt}\address{Universit\`a degli Studi Roma Tre,
Dip.to di Matematica e Fisica,
Via della Vasca Navale 84,
00146 Roma,
Italy}
\author{C. Motch}\address{Observatoire Astronomique Strasbourg,
11 rue de l'UniversitŽ,
67000 Strasbourg,
France}
\author{P. O'Brien}\address{University of Leicester,
Department of Physics \& Astronomy, 
University Rd,
Leicester LE1 7RH,
United Kingdom}
\author{G. Pareschi}\address{INAF-Osservatorio Astronomico di Brera,
Osservatorio astronomico di Brera,
Via Bianchi 46,
23807 Merate,
Italy}
\author{E. Pointecouteau}\address{$^{a}$Universit\'e de Toulouse; UPS-OMP; IRAP; Toulouse, France \& $^{b}$CNRS; Institut de Recherche en Astrophysique et Plan\'etologie; 9 Av. colonel Roche, BP 44346, F-31028 Toulouse cedex 4, France}
\author{G. Pratt}\address{CEA Saclay,
Service d'Astrophysique,
L'Orme des Merisiers Bat 709,
BP 2,
91191 Gif-sur-Yvette Cedex,
France}
\author{G. Rauw}\address{Institut d'Astrophysique-G\'eophysique,
Universit\'e de Li\`ege,
IAGL,
17 all\'ee du 6 Ao\^ut (B\^at B5c),
Sart Tilman,
4000 Li\`ege,
Belgium}
\author{T. Reiprich}\address{Argelander Institute for Astronomy,
Bonn University,
Auf dem H\"ugel 71,
53121 Bonn,
Germany}
\author{J. Sanders}\address{Max-Planck-Institut fur extraterrestrische Physik, P.O. Box 1312, Giessenbachstrasse, 85741 Garching, Germany}
\author{S. Sciortino}\address{INAF,
Osservatorio Astronomico di Palermo G.S. Vaiana,
Piazza del Parlamento 1,
90134 Palermo,
Italy}
\author{R. Willingale}\address{University of Leicester,
Department of Physics \& Astronomy, 
University Rd,
Leicester LE1 7RH,
United Kingdom}
\author{J. Wilms}\address{University of Erlangen-Nuremberg,
Dr. Karl Remeis-Sternwarte and ECAP,
Sternwartstr. 7,
96049 Bamberg,
Germany}
\author{on behalf of the The Hot and Energetic Universe white paper contributors and the Athena+ supporters}\address{Full lists can be accessed at \url{http://www.the-athena-x-ray-observatory.eu}}

\setcounter{page}{237}


\maketitle

\newpage
\begin{abstract}
The Advanced Telescope for High-energy Astrophysics ({\it Athena+}) is being proposed to ESA as the L2 mission (for a launch in 2028) and is specifically designed to answer two of the most pressing questions for astrophysics in the forthcoming decade:  How did ordinary matter assemble into the large scale structures we see today? and how do black holes grow and shape the Universe? For addressing these two issues, {\it Athena+} will provide transformational capabilities in terms of angular resolution, effective area, spectral resolution, grasp, that will make it  the most powerful X-ray observatory ever flown. Such an observatory, when opened to the astronomical community, will be used for virtually all classes of astrophysical objects, from high-z gamma-ray bursts to the closest planets in our solar neighborhood. In this paper, we briefly review the core science objectives of {\it Athena+}, present the science requirements and the foreseen implementation of the mission, and illustrate its transformational capabilities compared to existing facilities. 
\end{abstract}

\begin{keywords}
Accretion, accretion disks --
Equation of state --
Black hole physics --
Techniques: high angular resolution --
Techniques: imaging spectroscopy --
Techniques: spectroscopic --
Telescopes Surveys --
Stars: winds, outflows --
ISM: supernova remnants  --
Galaxy: center --
Galaxies: high-redshift --
(Galaxies:) intergalactic medium --
(Galaxies:) quasars: general --
(Galaxies:) quasars: supermassive black holes --
(Cosmology:) large-scale structure of Universe --
(Cosmology:) dark ages, reionization, first stars --
X-rays: binaries --
X-rays: galaxies --
X-rays: galaxies: clusters --
X-rays: general --
X-rays: ISM --
X-rays: stars 
\end{keywords}


\section{\ath\ science objectives: The Hot and Energetic Universe}
How did ordinary matter assemble into the large scale structures we see today? To answer this question we must trace the physical evolution of galaxy clusters and groups, as the most massive structures in the Universe, from their formation epoch to the present day. These structures grow over cosmic time by accretion of gas from the intergalactic medium, with the endpoint of their evolution being todayÕs massive clusters of galaxies, the largest bound structures in the Universe. Hot gas in clusters, groups and the intergalactic medium dominates the baryonic content of the local Universe, so understanding how this component forms and evolves is a crucial goal \citep{athwp}. 
\begin{figure}[ht!]
 \centering
 \includegraphics[width=0.65\textwidth,clip]{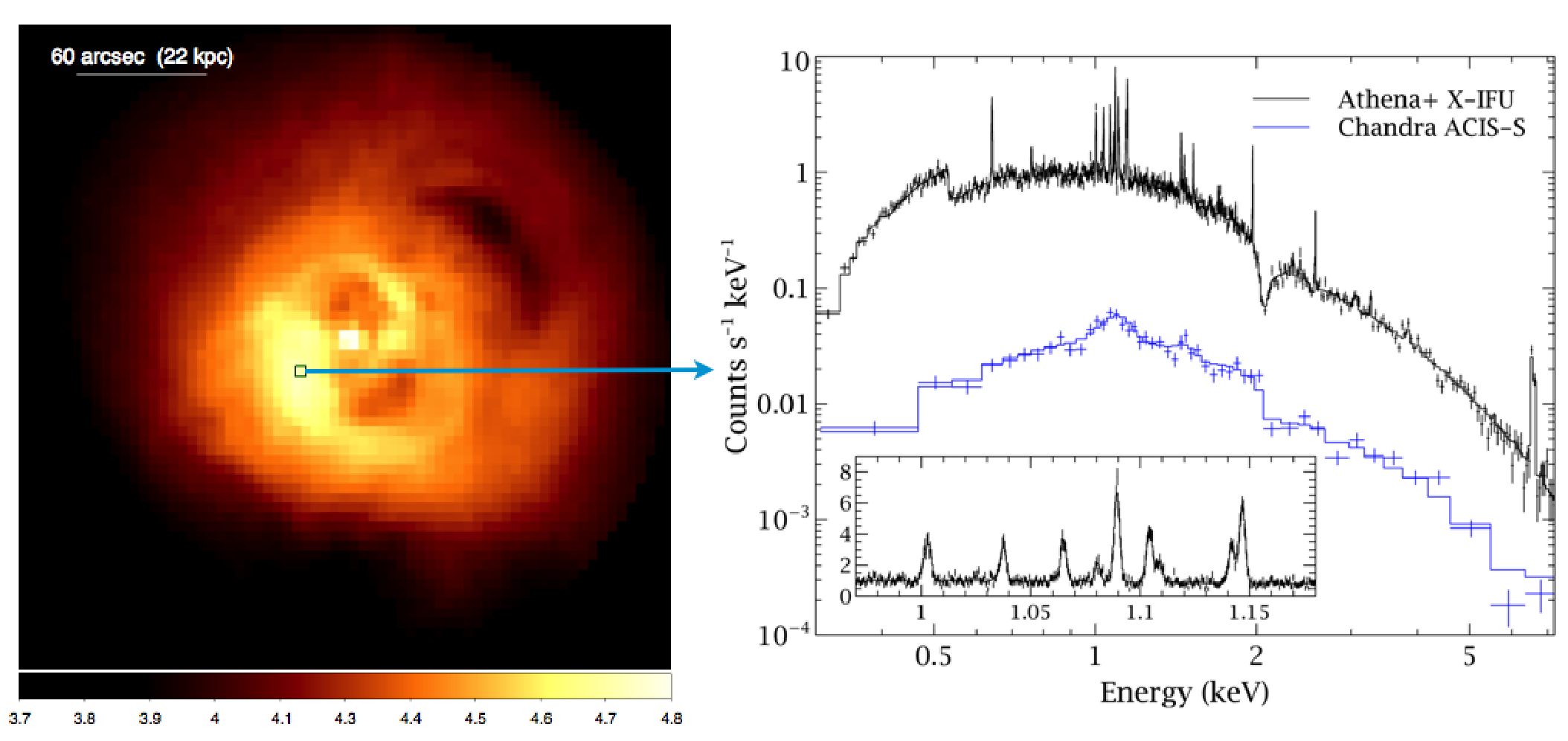}      
  \caption{Simulated \ath\ observations of the Perseus cluster, highlighting the advanced capabilities for revealing the intricacies of the physical mechanisms at play. The left panel shows a simulated 50 ks X-IFU observation (0.5-7 keV), displayed on a log scale. The spectrum on the right is from the single $5''\times 5''$ region marked by the box, with the existing \chan\ ACIS spectrum for comparison. The inset shows the region around the iron L complex. With such observations velocity broadening is measured to 10--20 km s$^{-1}$, the temperature to 1.5\% and the metallicity to 3\% on scales $<10$ kpc in 20--30 nearby systems, and on $<50$ kpc scales in hundreds of clusters and groups. Such measurements will allow us to pinpoint the locations of jet energy dissipation, determine the total energy stored in bulk motions and weak shocks, and test models of AGN fueling so as to determine how feedback regulates hot gas cooling \citep[from][]{athsp03}.}
  \label{barret:fig1}
\end{figure}

While the framework for the growth of structures is set by the large scale dark matter distribution, processes of an astrophysical origin also have a major effect  \citep[][and references therein]{athsp01,athsp02,athsp03}. To understand them, it is necessary to measure the velocities, thermodynamics and chemical composition of the gas to quantify the importance of non-gravitational heating and turbulence in the structure assembly process \citep{athsp01,athsp02}. The temperature of the hot gas is such that it emits primarily in the X-ray band, but current and planned facilities do not provide sufficient collecting area and spectral resolution to settle the key issue of how ordinary matter forms the large scale structures that we see today. The key breakthrough is to enable spectroscopic observations of clusters beyond the local Universe, out to z=1 and beyond, and spatially resolved spectroscopy to map the physical parameters of bound baryonic structures \citep{athsp01,athsp02}. Technological advances in X-ray optics  and instrumentation can deliver simultaneously a factor 10 increase in both telescope throughput and spatial resolving power for high resolution spectroscopy \citep{athsp14,athsp15}, allowing the necessary physical diagnostics to be determined at cosmologically relevant distances for the first time.  At even larger scales, the locations and kinematics of the missing baryons predicted to reside in the warm hot phase of the intergalactic medium and providing a tracer of the large scale dark matter structures of the local Universe, can only be fully revealed via high resolution X-ray spectroscopy \citep{athsp04}.   

One of the critical processes shaping hot baryon evolution is energy input Ð commonly known as feedback Ð from supermassive black holes \citep{athsp03,athsp07} (see Fig. \ref{barret:fig1}). Remarkably, processes originating at the scale of the black hole event horizon seem able to influence structures on scales 10 orders of magnitude larger \citep{athsp07,athsp08}. This feedback is an essential ingredient of galaxy evolution models, but it is not well understood. X-ray observations are again the key to further progresses, revealing the mechanisms which generate and launch winds close to black holes and determining the coupling of the energy and matter flows on larger galactic and  cluster scales \citep{athsp06,athsp07,athsp08}. 

\begin{figure}[ht!]
 \centering
 \includegraphics[width=0.5\textwidth,clip]{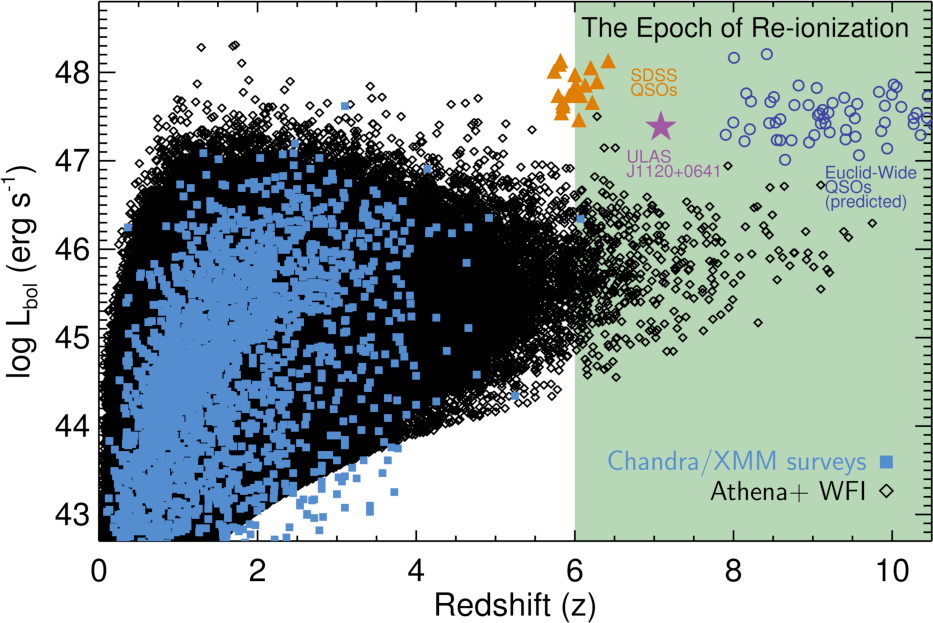}      
  \caption{Predictions for the redshifts and luminosities of 600,000 AGN that will be identified with a multilayered 1-year \ath\ WFI survey, compared to current \chan\ and \xmm\ surveys. \ath\ will discover hundreds of X-ray selected AGN in the epoch of reionization at $z>6$, sources which are around 2 orders of magnitude fainter than current optical and near-IR surveys (e.g. SDSS, UKIDSS), thus providing an essential complement to the luminous QSOs that Euclid will identify in the next decade \citep{roche2012mnras}, including in particular the most obscured ones missed by those surveys \citep[for a discussion see,][]{athsp05}.}
  \label{barret:fig2}
\end{figure}

The second key question to be addressed by \ath\ is that of how black holes grow and shape the Universe \citep{athwp}. The widespread importance of black hole feedback means that we cannot have a complete understanding of galaxies without tracking the growth of their central supermassive black holes through cosmic time. A key goal is to push the frontiers of black hole evolution to the redshifts where the first galaxies are forming, at $z=6-10$  \citep[][and references therein]{athsp05,athsp06}. X-ray emission is the most reliable and complete way of revealing accreting black holes in galaxies, but survey capabilities need to be improved by a factor $\sim 100$ over current facilities to reach these early epochs and perform a census of black hole growth \citep{athsp05} (see Fig. \ref{barret:fig2}). This requires a combination of high sensitivity, which in turn depends on large throughput and good angular resolution, and wide field of view. Again, the required technologies to provide this leap in wide field X-ray spectral imaging are now within our grasp \citep{athsp16}. The same high throughput needed to detect these early black holes will also yield the first detailed X-ray spectra of accreting black holes at the peak of galaxy growth at $z=1-4$, measurements which are impossible with current instrumentation. These spectra will show, for example, if the heavily obscured phase of black hole evolution is associated with the termination of star formation in galaxies via feedback \citep{athsp06}.

The Hot and Energetic Universe includes almost all known astrophysical objects, from the closest planets to the most distant quasars and gamma-ray bursts. The instrument suite required to achieve the science goals described above provides \ath\ with unprecedented observatory capabilities, enabling breakthrough observations to be performed for a wide range of objects, of interest to the whole astronomical community. This covers Solar system bodies and exoplanets, stars, compact objects , supernova remnants, the interstellar medium, and luminous extragalactic transients, such as gamma-ray bursts. Key topics to be addressed include: i) Establish how planetary magnetospheres and exospheres, and comets, respond to the interaction with the solar wind, in a global way that in situ observations cannot offer  \citep{athsp09}, ii) assess the mass loss rates of high velocity chemically-enriched material from massive stars to understand the role they play in the feedback processes on Galactic scales  \citep{athsp10}, iii) discover how mass loss from disk winds influences the binary evolution and impact the interstellar medium \citep{athsp11}, iv) understand the physics of core collapse and type Ia supernova remnants and determine the chemical composition of the hot and cold gas of the interstellar medium, as a tracer of stellar activity in our and other galaxies  \citep{athsp12}, v) distinguish Population III from Population II star forming regions as gamma-ray burst progenitors \citep{athsp13}. 

\begin{figure}[ht!]
 \centering
 \includegraphics[width=0.41\textwidth,clip]{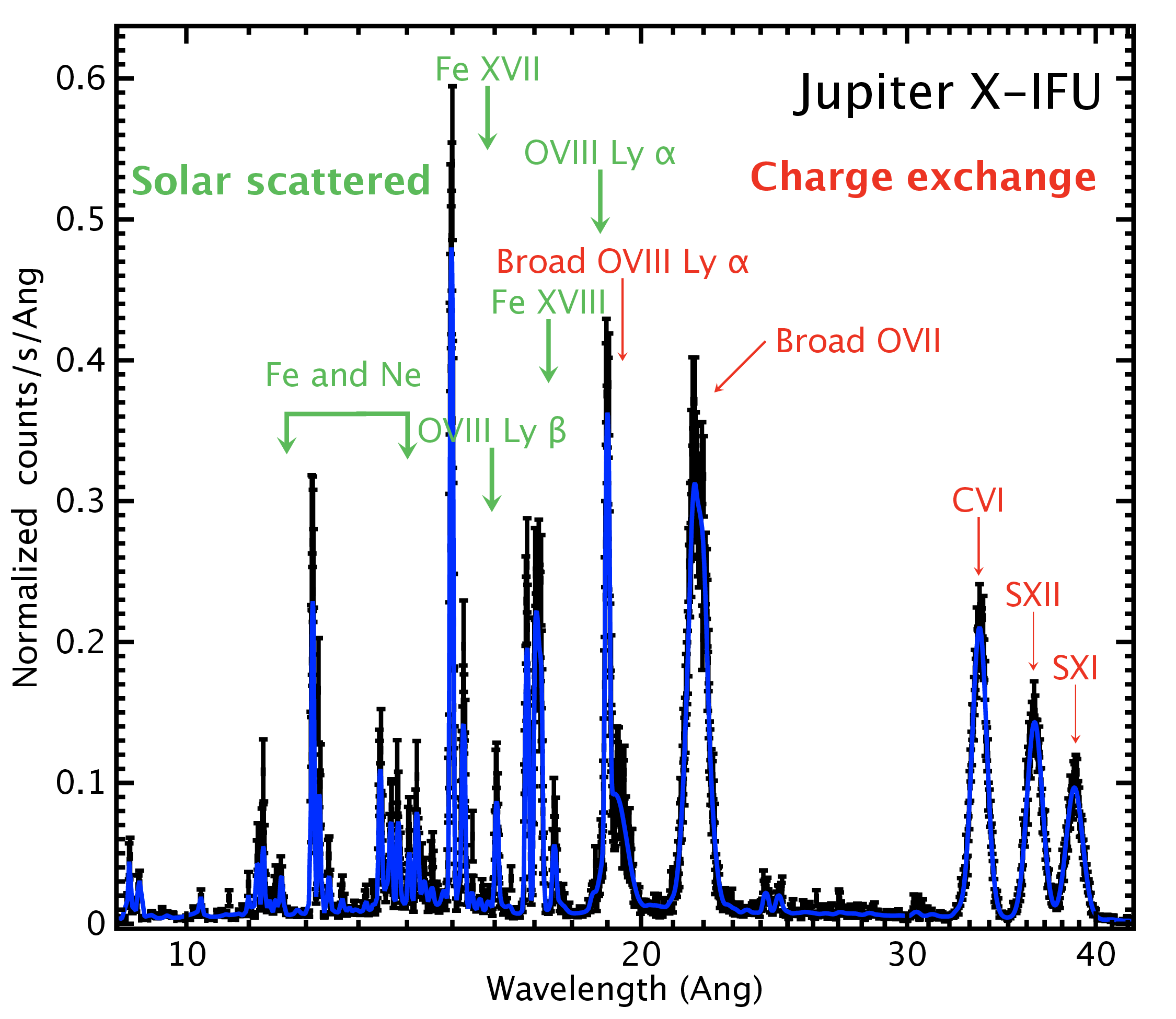}\includegraphics[width=0.41\textwidth,clip]{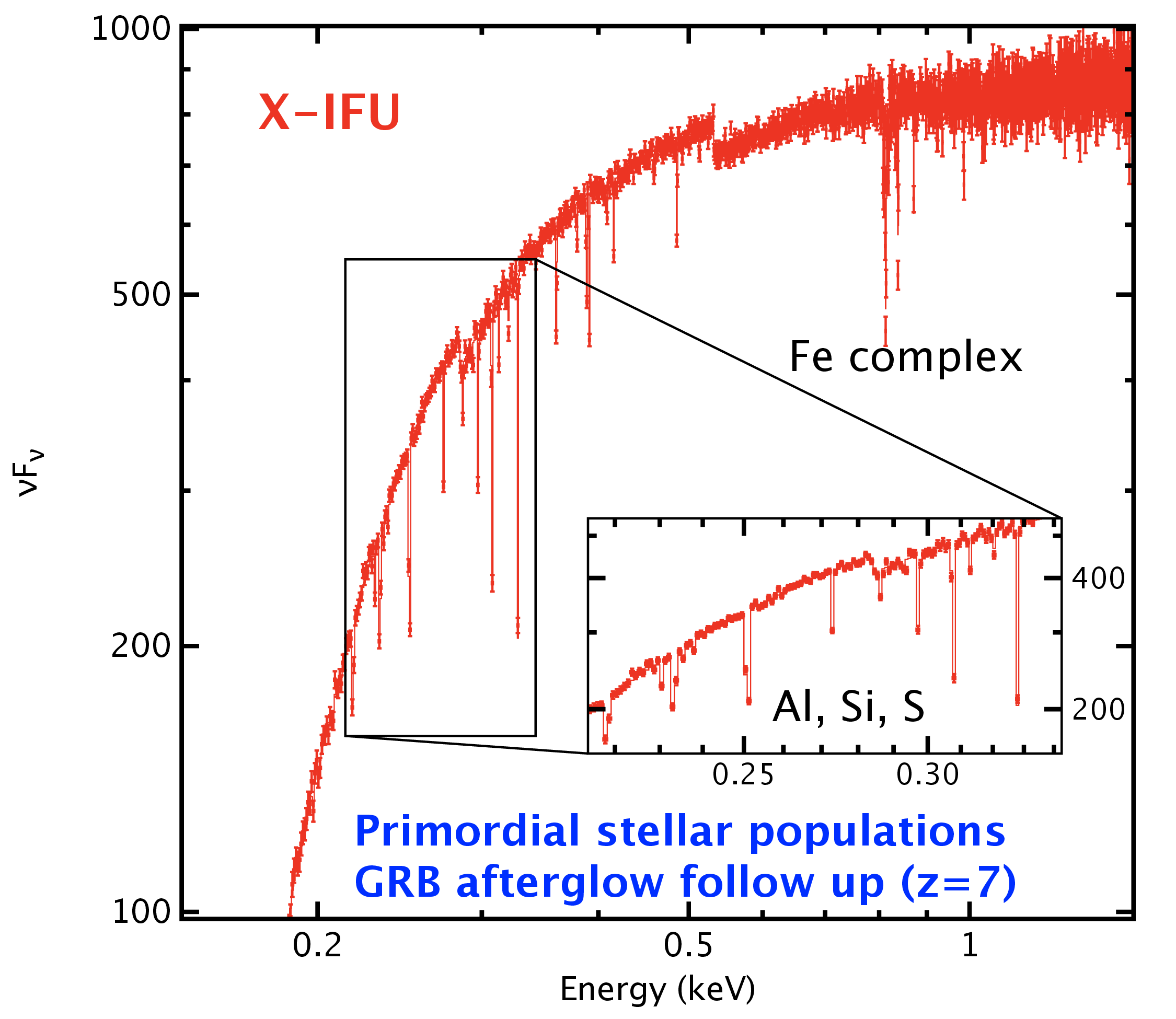} 
  \caption{{\bf Left:} Simulated \ath\ X-IFU spectrum of Jupiter for an exposure time of 20 ks: the wavelength coverage extends to the 30 -- 40 \AA~band allowing to resolve the C/S ambiguity  \cite[from][]{athsp09}. {\bf Right:} A simulated X-IFU X-ray spectrum of a GRB afterglow at $z=7$, showing the capability of \ath\ in tracing the primordial stellar populations. This medium bright afterglow (fluence=$0.4\times10^{-6}$ erg cm$^{-2}$) is characterized by deep narrow resonant lines of Fe, Si, S, Ar, Mg, from the ionized gas in the environment of the GRB. An effective column density of $2\times10^{22}$ cm$^{-2}$ has been adopted. The abundance pattern measured by \ath\ can distinguish Population III from Population II star forming regions \cite[from][]{athsp13}.}
  \label{barret:fig3}
\end{figure}
\section{The \ath\ mission concept}
\ath\ in an X-ray observatory-class mission delivering a transformational leap in high-energy observational capabilities.  A lightweight X-ray telescope based on ESA's Silicon Pore Optics (SPO) technology provides large effective area with excellent angular resolution, to be combined with state-of-the-art instrumentation for spatially resolved high resolution X-ray spectroscopy \cite[provided by the X-ray Integral Field Unit, X-IFU,][]{athsp15} and wide field X-ray imaging \cite[provided by the Wide Field Imager, WFI, ][]{athsp16}. \ath\ will thus deliver superior wide field X-ray imaging spectroscopy and timing capabilities, far beyond those of any existing or approved future facilities. Mapping the dynamics and chemical composition of hot gas in diffuse sources requires high spectral resolution imaging (2.5 eV resolution) with low background; this also optimizes the sensitivity for weak absorption and emission features needed for WHIM studies or for faint point source characterisation. An angular resolution of 5 arcsec is required to disentangle small structures of clusters and groups and, in combination with a large area, provides high resolution spectra, even for faint sources. This angular resolution, combined with the mirror effective area and large field of view (40 arcmin) of the WFI provides the deep survey area and detection sensitivity (limiting flux of $10^{-17}$ erg cm$^{-2}$ s$^{-1}$ $0.5-2$ keV band) required to detect AGN at $z>6$ within a reasonable survey time. The science requirements and enabling technologies for \ath\ are summarized in Table 1.
\begin{table}[htbp]
\small
   \centering
   \begin{tabular}{p{5cm}p{5cm}p{5cm}} 
\hline
\hline
Parameter & Requirements & Enabling technology/comments  \\
\hline
Effective Area	 & 2 m$^2$ @ 1 keV (goal 2.5 m$^2$) &  Silicon Pore Optics developed by\\
 &0.25 m$^2$ @ 6 keV (goal 0.3 m$^2$)	  & ESA. Single telescope: 3 m outer diameter, 12 m fixed focal length. \\
\hline
Angular Resolution	 & $5''$ (goal 3'') on-axis & Detailed analysis of error budget \\
& 10'' at 25' radius 	 & confirms that a performance of 5'' HEW is feasible. \\
\hline

Energy Range	  & 0.3-12 keV & 	Grazing incidence optics. \\
\hline

Instrument Field of View & 	Wide-Field Imager: (WFI): 40' (goal 50')& Large area DEPFET Active Pixel Sensors.  \\
	& X-ray Integral Field Unit: (X-IFU): 5' (goal 7') & 	Large array of multiplexed Transition Edge Sensors (TES) with 250 $\mu$m pixels. \\
\hline

Spectral Resolution	 & WFI: $<150$ eV @ 6 keV	 & Large area DEPFET Active Pixel Sensors.\\
		 &X-IFU: 2.5 eV @ 6 keV (goal 1.5 eV @ 1 keV)	 & Inner array (10''x10'') optimized for goal resolution at low energy (50 $\mu$m pixels). \\
\hline

Count Rate Capability	 & $> 1$ Crab  (WFI)	 & Central chip for high count rates without pile-up and with micro-second time resolution.\\
	 & 10 mCrab, point source (X-IFU) & Filters and beam diffuser enable \\
 & 1 Crab (30\% throughput) & 	higher count rate capability with reduced spectral resolution.  \\
\hline
TOO Response & 	4 hours (goal 2 hours) for 50\% of time & 	Slew times $<2$ hours feasible; total response time dependent on ground system issues. \\
\hline
\hline

   \end{tabular}
   \caption{Key parameters and requirements of the \ath\ mission. The enabling technology is indicated.}
   \label{tab:booktabs}
\end{table}

\begin{figure}[ht!]
 \centering
\includegraphics[width=0.333\textwidth,clip]{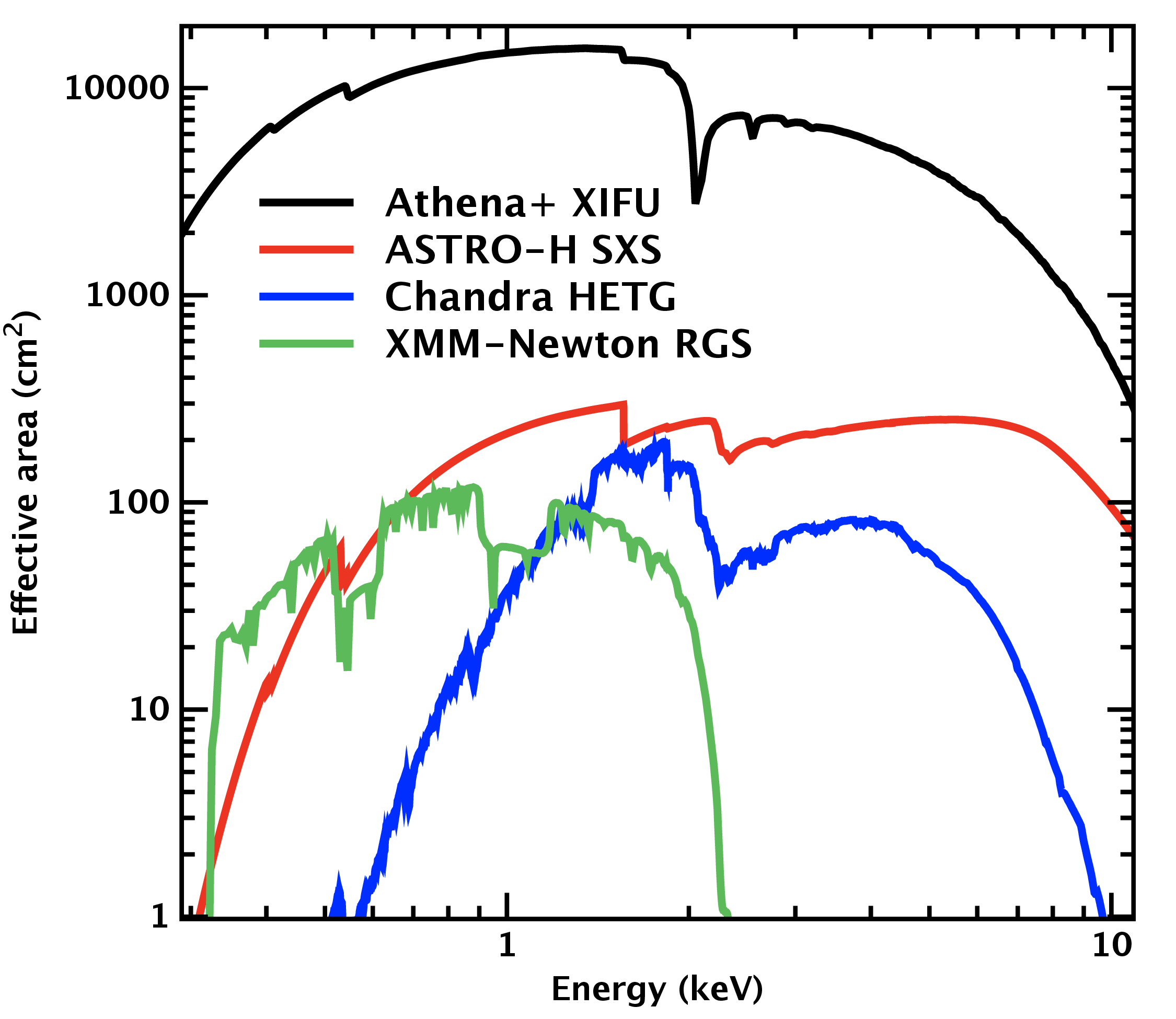}\includegraphics[width=0.333\textwidth,clip]{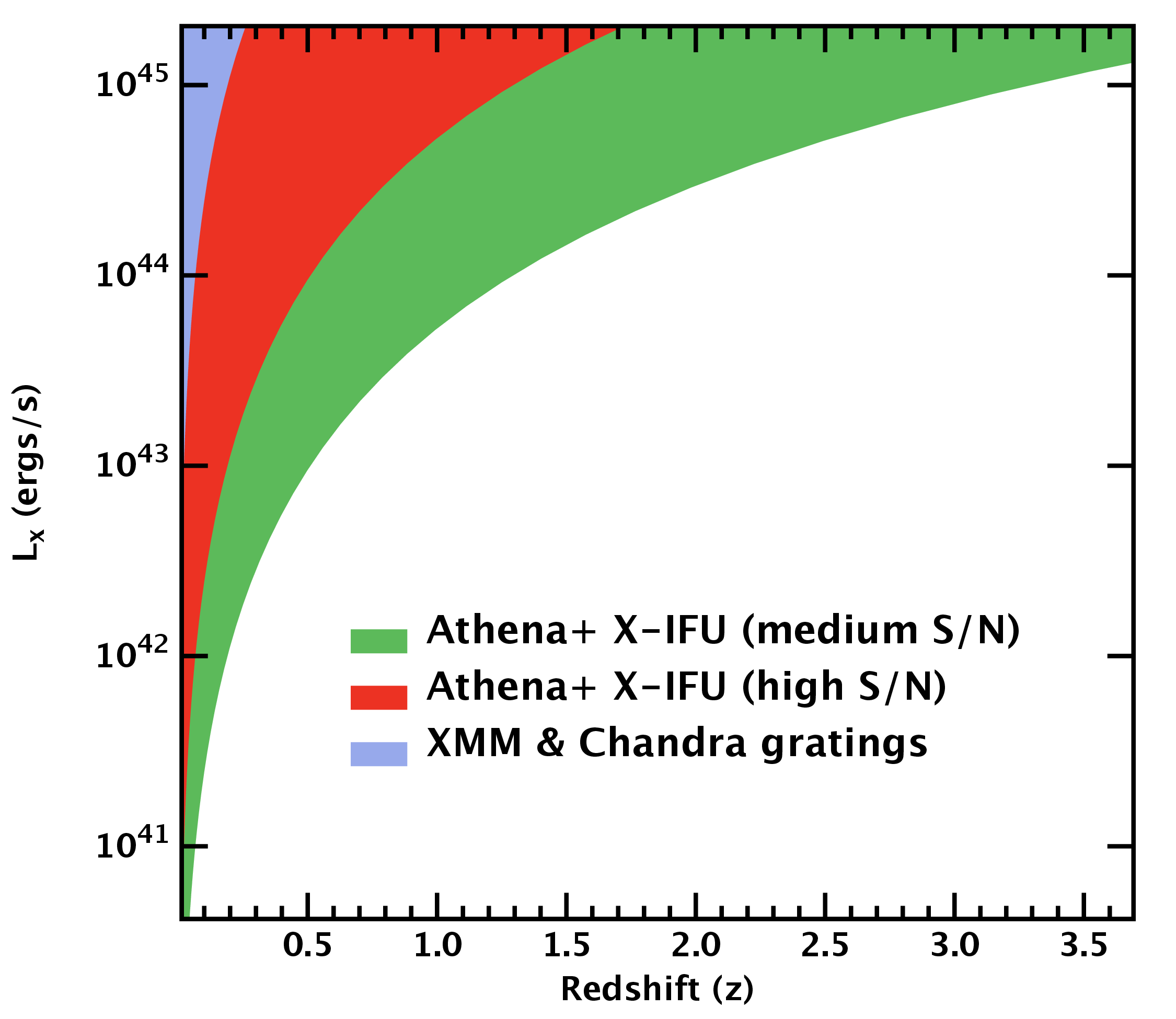}\includegraphics[width=0.333\textwidth,clip]{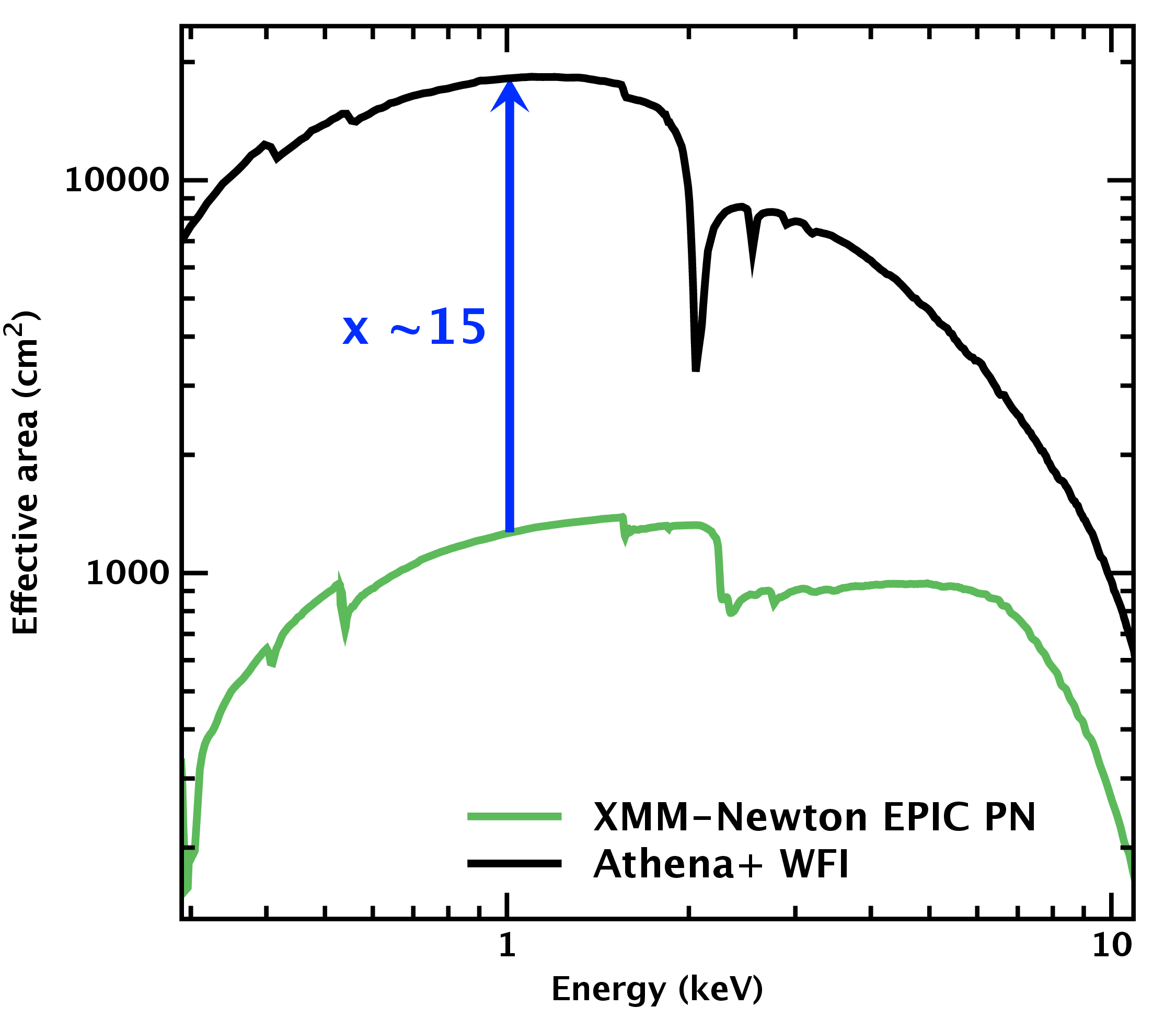}
\includegraphics[width=0.333\textwidth,clip]{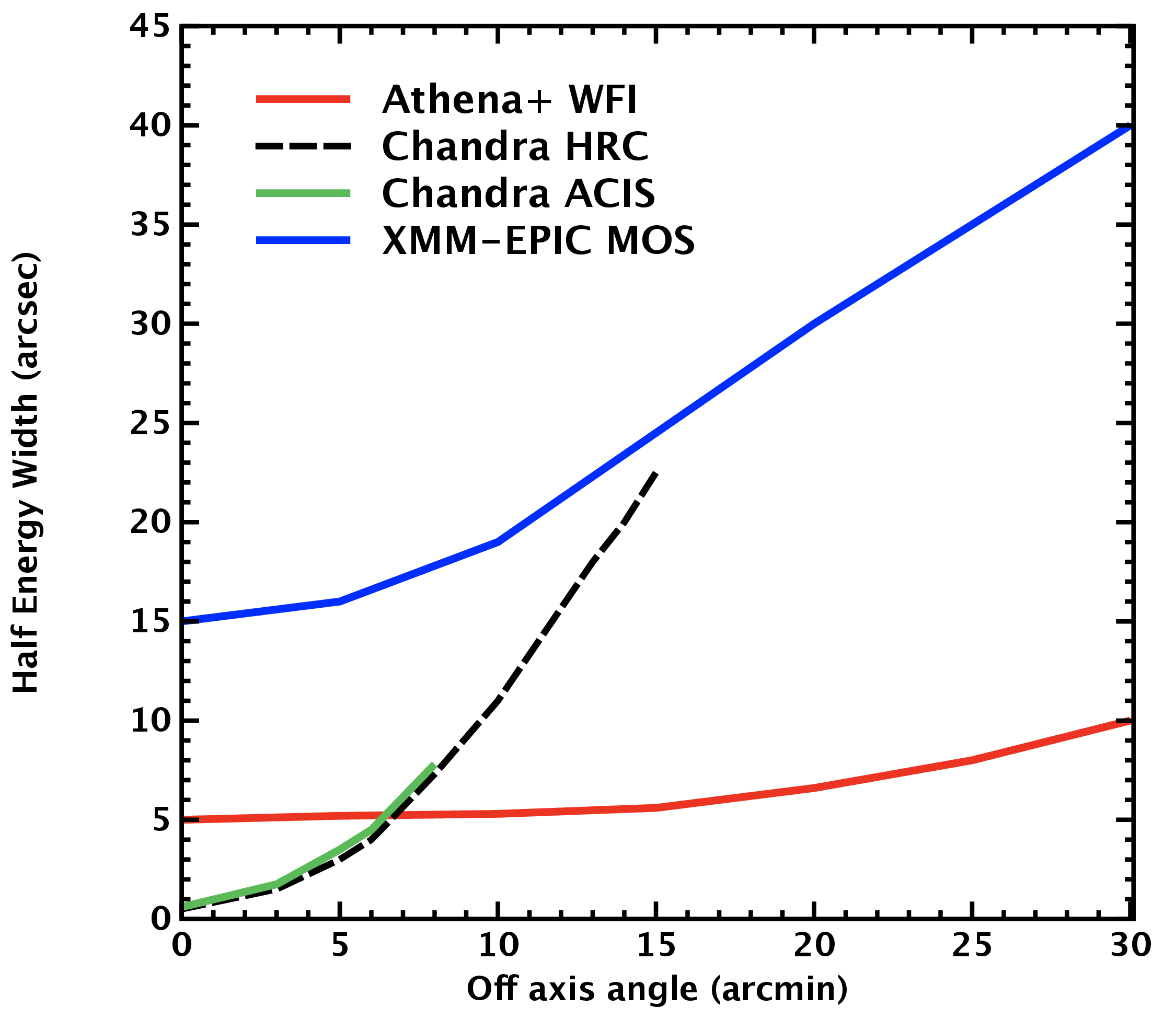}\includegraphics[width=0.333\textwidth,clip]{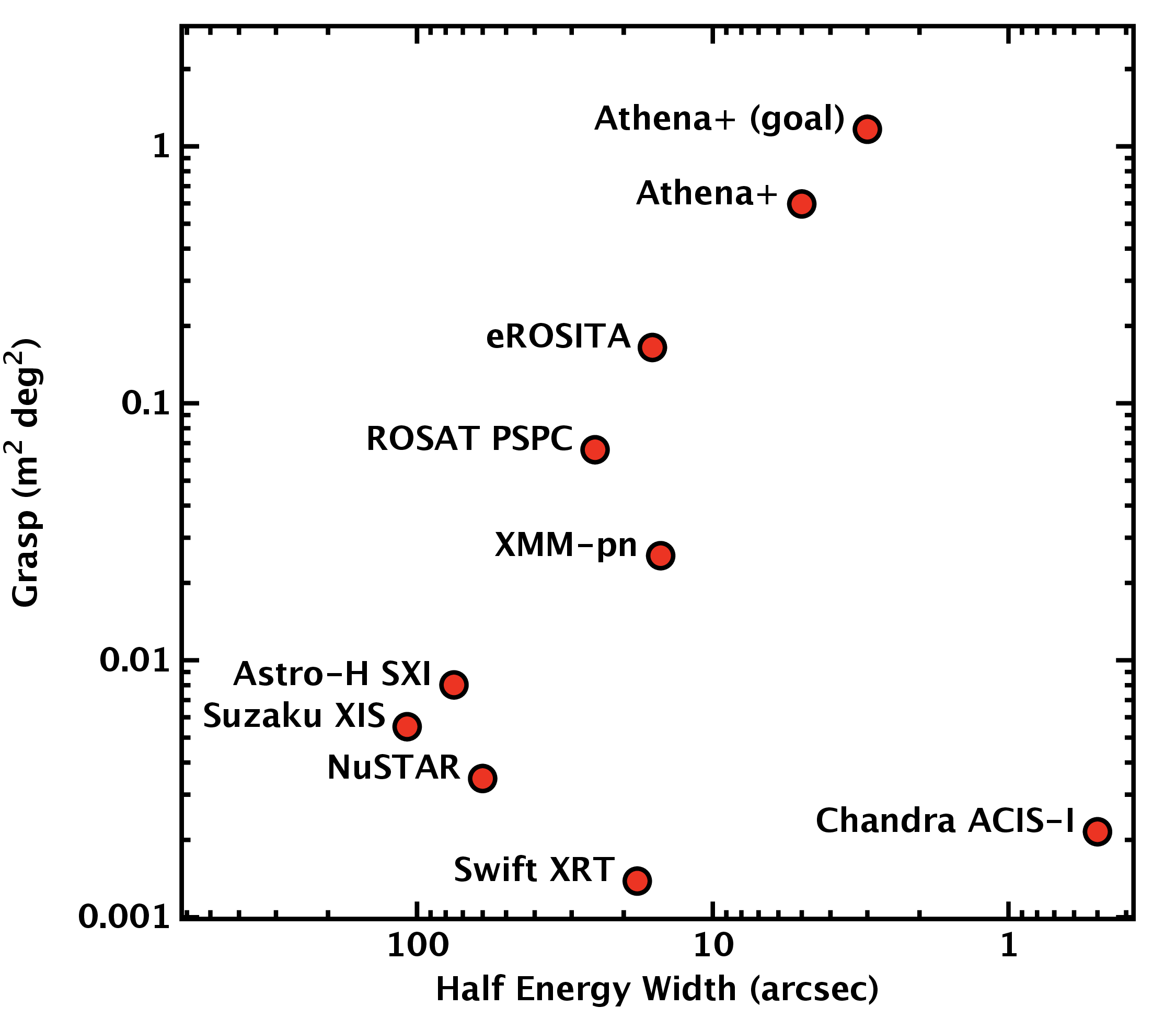}\includegraphics[width=0.333\textwidth,clip]{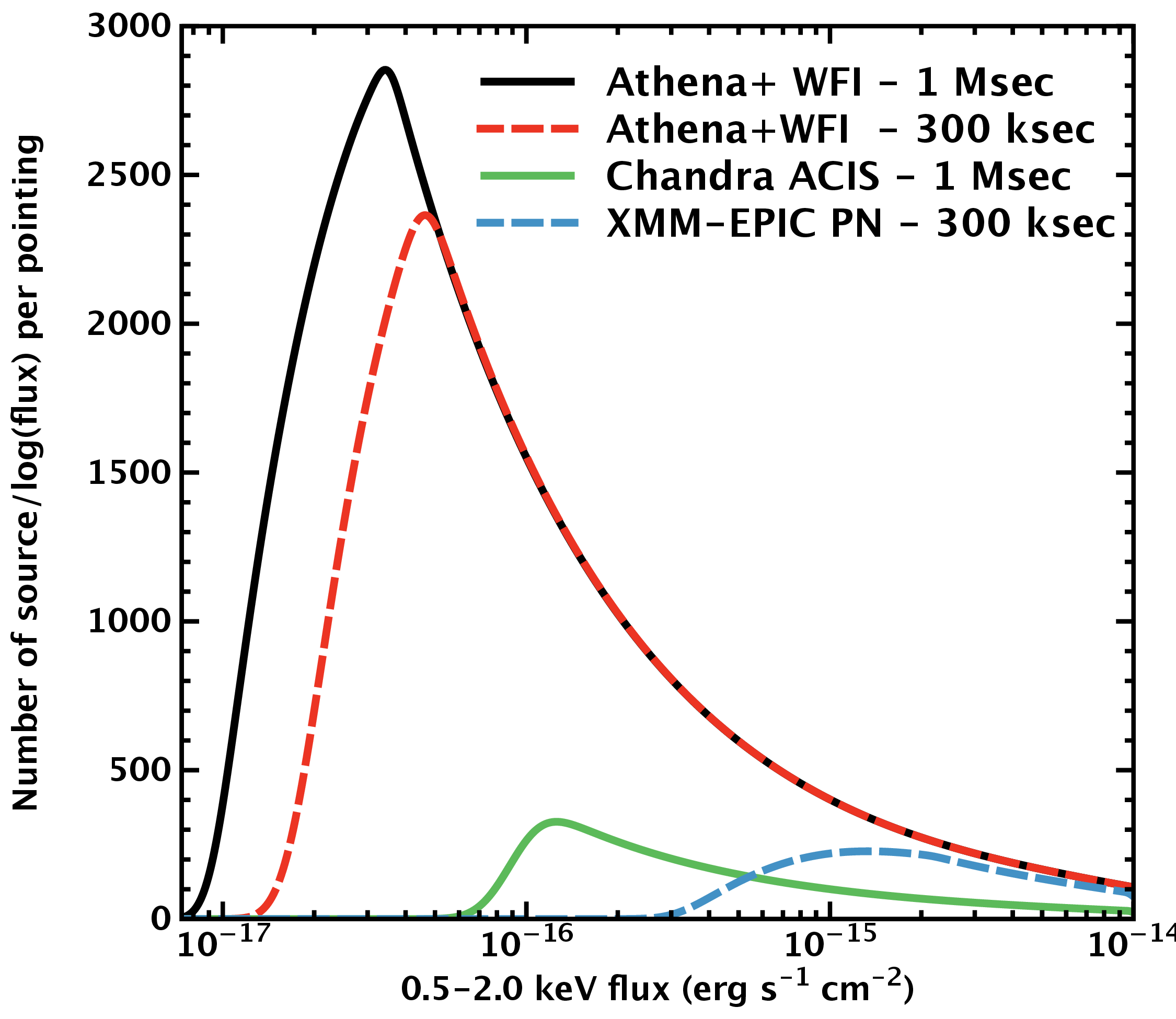} 
  \caption{ {\bf Top left:} Effective area curves for high resolution X-ray spectrometers, operational and planned. {\bf Top center:} $L_{x}-z$ curve for high resolution X-ray spectroscopy, compared to current X-ray gratings. {\bf Top right:} Effective area curves for spectral-timing-imaging observations with the \xmm\ PN camera and the \ath\ wide field imager. {\bf Bottom left:} Half Energy Width (arcsec) (PSF variation across the field of view) comparison between existing X-ray imagers and the \ath\ wide-field imager. {\bf Bottom center:} Grasp of previous, operational and planned missions as a function of angular resolution. Grasp is defined as the product of effective area at 1 keV (10 keV for {\it NuSTAR}) and the instrument field of view. {\bf Bottom right:} Number of sources per logarithmic flux interval expected in single \ath\ WFI pointings at high Galactic latitudes compared to \chan\ and \xmm. }
  \label{barret:fig4}
\end{figure}

The strawman \ath\ payload comprises three key elements:

\begin{itemize}
\item A single X-ray telescope with a focal length of 12 meters and an unprecedented effective area (2 m$^2$ at 1 keV). The X-ray telescope employs Silicon Pore Optics (SPO), an innovative technology that has been pioneered in Europe over the last decade mostly with ESA support. SPO is a highly modular concept, based on a set of compact individual mirror modules, which has an excellent effective area-to-mass ratio and can achieve high angular resolution ($<5''$) and a flat response across the field of view \citep{athsp14} (see Fig. \ref{barret:fig4}).
\item The X-ray Integral Field Unit (X-IFU), an advanced actively shielded X-ray microcalorimeter spectrometer for high-resolution imaging, utilizing Transition Edge Sensors cooled to 50 mK \citep{athsp15}.
\item The Wide Field  Imager (WFI), a Silicon Active Pixel Sensor camera with a large field of view, high count-rate capability and moderate resolution spectroscopic capability \citep{athsp16}.
\end{itemize}

\ath\ is based on a conventional design retaining much heritage from \xmm. Considerations of observing efficiency and thermal stability favour an L2 orbit reached by Ariane V.  \ath\ will be an observatory whose program will be largely driven by calls for proposals from the scientific community, but may be complemented by key programs for science goals requiring large time investments. A nominal mission lifetime of 5 years would allow the core science goals set out in the White Paper to be achieved, while preserving a large fraction of the available time for broad based science programs \citep{athwp}.

With such a mission configuration \ath\ will truly provide transformational capabilities, as illustrated in Fig. \ref{barret:fig4}. It will provide an improvement factor $>100$ in high spectral resolution throughput (e.g. compared to {\it ASTRO-H} microcalorimeter and \xmm\ gratings). It will open high-resolution X-ray spectroscopy to the high redshift Universe (up to $z\sim3$ without mentioning gamma-ray bursts), while it is currently limited to redshifts less than $\sim 0.3$. It will provide a factor of more than 10 in bare throughput (i.e. area on axis)  for spectral-timing-imaging observations compared to the \xmm\ PN camera. Thanks to the combination of effective area, field of view, angular resolution (and its flatness across the field of view), \ath\ will deliver an improvement factor larger than $100$ in survey speed compared to \chan\ (see Fig. \ref{barret:fig4}, lower panels).

\section{Conclusions}
\ath\ is designed to tackle two of the most pressing questions of modern astrophysics: How did ordinary matter assemble into the large scale structures we see today? and how do black holes grow and shape the Universe? \ath\ provides the necessary angular resolution, spectral resolution, throughput, detection sensitivity, and survey grasp to answer these two questions. The technologies for \ath\ are mature, being based on much previous heritage and major technology developments in Europe. \ath\, as an X-ray observatory, will open up a vast discovery space leading to completely new areas of scientific investigation, continuing the legacy of discovery that has characterized X-ray astronomy since its inception. The implementation of \ath\ for launch in 2028 will guarantee a transformation in our understanding of The Hot and Energetic Universe, and establish European leadership in high-energy astrophysics for the foreseeable future.
\begin{acknowledgements}
We gratefully acknowledge the comments and inputs of the White Paper Review Team: M. Arnaud, J. Bregman, F. Combes, R. Kennicutt, R. Maiolino, R. Mushotzky, T. Ohashi, K. Pounds, C. Reynolds, H. R\"{o}ttgering, M. Rowan-Robinson, C. Turon and G. Zamorani. It is also a pleasure to thank the 1178 \ath\ supporters, recorded as of \today.\footnote{All information related to \ath, including the list of supporters can be found on the \ath\ web site: \url{http://www.the-athena-x-ray-observatory.eu}.}
\end{acknowledgements}


%
\end{document}